\documentclass[%
 aip,
 amsmath,amssymb,
 reprint,%
]{revtex4-2}

\usepackage{graphicx}
\usepackage{overpic}
\usepackage{dcolumn}
\usepackage{bm}

\usepackage[utf8]{inputenc}
\usepackage[T1]{fontenc}
\usepackage{mathptmx}
\usepackage{etoolbox}
\usepackage{xcolor}
\usepackage{hyperref}

\newcommand{\xx}{\mathbf{x}}

\newcommand{\vv}{\mathbf{v}}

\newcommand{\BB}{\mathbf{B}}
\newcommand{\RR}{\mathbb{R}}
\newcommand{\SL}{\mathrm{SL}_2(\mathbb{R})}
\newcommand{\Tr}{\mathrm{Tr}}

\newcommand{\MM}{\mathsf{M}}

\newcommand{\Ma}{\mathcal{M}}

\graphicspath{{fig/}}

\makeatletter
\def\@email#1#2{%
 \endgroup
 \patchcmd{\titleblock@produce}
  {\frontmatter@RRAPformat}
  {\frontmatter@RRAPformat{\produce@RRAP{*#1\href{mailto:#2}{#2}}}\frontmatter@RRAPformat}
  {}{}
}%
\makeatother
\begin{document}

\title{Surfaces with Klein bottle topology occur in fusion reactor fields}
\author{C. B. Smiet}
\affiliation{Ecole Polytechnique Fédérale de Lausanne (EPFL), Swiss Plasma Center (SPC), CH-1015 Lausanne, Switzerland}
 \email{christopher.smiet@epfl.ch.}

\date{\today}

\begin{abstract}
Magnetic confinement fusion devices, such as tokamaks and stellarators, are designed such that magnetic field lines lie in magnetic surfaces that form a foliation of nested genus-1 tori. 
The Poincar\'e-Hopf index theorem implies that only surfaces with genus 1 allow for vector fields that lie tangent to the surface and are smooth and nowhere-vanishing. 
It is often implicitly assumed that magnetic surfaces are always toroidal. 
This paper we show that surfaces with the topology of an immersed Klein bottle (the only other genus-1 compact surface) can also occur in fusion reactor fields. 
  These surfaces occur around fixed points of the Poincar\'e map that are reflection-hyperbolic, and are spanned by the field lines that asymptotically approach and depart from this closed line.   
Configurations in which this occurs appear in the stellarator database QUASR, and have been described in literature in the context of abnormal satwooth crashes. 
\end{abstract}

\maketitle

This paper does not solve any key problems on the path to realizing fusion, nor does it claim that its conclusions are important. 
It exists purely because of the aesthetically pleasing nature of the novel observations it contains, highlighting profound connections between pure mathematics, applied mathematics, and physics. 

It has often been lamented by plasma physicists that if only we lived in a universe with four spatial dimensions, fusion would be easy to solve.\footnote{
This lamentation comes mostly from myself, though I have seen S. R. Hudson nod bemusedly when discussing this with him.}
The straight Clifford torus would eliminate all curvature, and thus its associated drifts, and interesting shapes such as Klein bottle fields would be realizable. 
One must forgive the mathematicians' folly in attempting to rise out of our flatland, but any attempt to realize a fusion reactor must involve immersing it in our undeniably three-dimensional reality. 
We will show that despite this limitation, some magnetic surfaces with the topology of an immersed Klein bottle can occur in fusion reactor fields.

Magnetic confinement fusion reactors confine a plasma by creating a strong magnetic field. 
The charged particles of the plasma, to first order, follow these field lines, and are thus confined. 
The divergence-free nature of the magnetic field, enshrined in the Maxwell equation $\nabla\cdot\BB=0$, implies that magnetic field lines cannot start or end. 
Instead, field lines fall into one of four categories: 
\begin{enumerate}
    \item \textbf{closed curve:} The field line closes on itself after a finite distance.
    \item \textbf{surface:} The field line is confined to a two-dimensional surface. 
    \item \textbf{chaotic:} The field line fills a volume of space. 
    \item \textbf{unbounded:} the field line goes to infinity in a way that defies classification into the above categories. 
\end{enumerate}


The field lines in category 2 lie on a surface, and a \emph{magnetic surface} is a compact surface to which the field is everywhere tangent. 
Sometimes a single field line can fill that surface, for example if the surface is a simple torus where field lines on a cross-section returns to the same cross-section with on average an irrational rotation. 
The average rotation is called the rotational transform $\imath$ of the surface, In the context of Tokamak reactors it's inverse $q=1/\imath$ is often used. 
Because the rotational transform is irrational, a field line on  on this surface returns to a new point on each application of the map, and never closes on itself. 
A surface with rational rotational transform of $n/m$ contains only closed curves that close on themselves after $m$ circuits.
Magnetic fields consisting of a foliation of nested simple toroidal magnetic surfaces are the basis of magnetic confinement fusion in tokamak and stellarator reactors.  
These fields are generated by an external set of coils and/or currents internal to the plasma and the field does not vanish anywhere in the plasma region. 

The Poincar\'e-Hopf index theorem (whose special case is the well-known 'hairy ball theorem') relates the zeros of a vector field tangent to a manifold $\mathcal{M}$ to the Euler characteristic $\chi(\mathcal{M})$. 
A consequence is that only $\chi=0$ surfaces can sustain vector fields that do not vanish anywhere.
The Euler characteristic, which for orientable surfaces is related to the genus $g$ (number of 'holes' in a surface) through $\chi = 2 - 2g$, is zero for the torus.

A surface to which the field is everywhere tangent trivially defines a vector field tangent to that surface (the magnetic field).
If the magnetic field is nowhere-vanishing in a region, then obviously it cannot vanish on a magnetic surface in that region. 
Therefore it can only lie tangent to surfaces with Euler characteristic $\chi=0$. (see also~\cite{bolsinov2004integrable}). 
Other geometries such as spherical surfaces or higher-genus tori, must necessarily contain points where the field vanishes. 

\begin{figure}[h]
  \centering
  \begin{minipage}{0.45\linewidth}
    \centering
    \includegraphics[width=\linewidth]{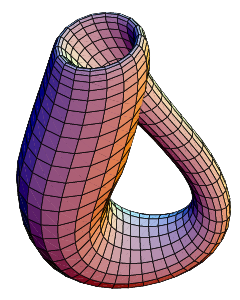}
  \end{minipage}
  \begin{minipage}{0.45\linewidth}
    \centering
    \includegraphics[width=\linewidth]{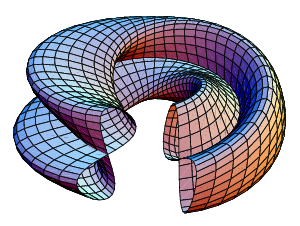}
  \end{minipage}
  \hfill
  \caption{Two immersions of the Klein bottle in $\RR^3$. (left) most common immersion. (right) lemniscate Klein bottle, the surface swept out by a lemniscate (sideways figure-8) that maps to itself with half-integer rotation. Images in public domain. Source: Wikimedia commons\footnote{\url{https://commons.wikimedia.org/wiki/File:KleinBottle-Figure8-02.png}, \url{https://commons.wikimedia.org/wiki/File:KleinBottle-01.png}} }\label{fig:Klein_bottle}
\end{figure}

There exist only two compact surfaces with genus 1, the first is the torus already discussed, the second is the Klein bottle.
There exist several immersions of the Klein bottle~\cite{franzoni2012Klein}, two of which are shown in figure~\ref{fig:Klein_bottle}. 
The 'usual' immersion (left) is the most common, described by Klein~\cite{Klein1882uber} as:
\begin{quote}
Man kann sich von denselben ein Bild machen , indem man etwa ein Stück eines Kautschukschlauches umstülpt und nun so sich selbst durchdringen lässt , dass bei Zusammenbiegung der Enden die Aussenseite mit der Innenseite zusammen- kommt . 
\end{quote}
(One can imagine taking a rubber hose, folding inside out and  making it penetrate itself, and connecting the ends in such a fashion that the outside meets the inside).
Another immersion is the lemniscate Klein bottle shown in figure~\ref{fig:Klein_bottle} (right), which was first described in 1967 by~\citet{banchoff1976minimal}. This form is obtained by sweeping a lemniscate ($\infty$-symbol) rotating it an integer and a half number of times before it is brought back to itself. 

It is the lemniscate Klein bottle that can appear in fusion reactor magnetic fields and in the trajectories of dynamical systems in three dimensions. 
The self-intersection point of the lemniscate forms a single field line that closes on itself, which is of reflection-hyperbolic nature~\cite{lichtenberg1992chaotic}.

In this paper we show examples of lemniscate immersed Klein bottles in fusion relevant fields.
Though fields in which this magnetic geometry occurs have been discussed in literature, their identification as an immersed Klein bottle has not been made. 
In section \ref{sec:maths} we describe the structure of the magnetic field around closed field lines. 
In section~\ref{sec:index} we discus index theorems and show that the 'usual' Klein bottle immersion cannot occur in magnetic fields. 
In section~\ref{sec:realisation} we discuss realizations of magnetic fields with immersed Klein bottle topology in fusion reactors. 
In section~\ref{sec:conclusions} we conclude and discuss trajectories with Klein bottle topology in dynamical systems literature and in other branches of physics. 

\section{The field line map around a closed magnetic field line \label{sec:maths}}
Magnetic field line flow in three dimensions is, under certain conditions, mathematically equivalent to trajectory flow in Hamiltonian dynamical system~\cite{kerst1962influence, morrison2000magnetic, duignan2024global}.
In order to analyze the trajectories in such a system, or the structure of the magnetic field, one constructs a Poincar\'e section. 
This is done by taking a plane transverse to the flow of the dynamics (magnetic field), and recording the locations at which orbits (magnetic field line) intersects that plane. 

We will use a cylindrical coordinate system $(R, \phi, Z)$ and choose as our  section the half-plane with $\phi=\text{const}$. 
The Poincar\'e map $f$ is the map that takes a point $\xx = (R, Z)$ in our section, and maps it to the point $f(\xx)$. 
A \emph{fixed point} is a point where $f(\xx)=\xx$. 
A field line that closes after $n$ applications is a fixed point of $f^n$. 

We can linearise $f$ around a fixed point such that
\begin{equation}\label{eq:linmap}
  f(\xx + \delta\xx) = f(\xx) + \MM\delta\xx
\end{equation}
where $\MM := \MM_{ij} = \partial_j f_i$ is the matrix of partial derivatives. 

At a fixed point $\nabla\cdot\BB=0$ implies that the mapping is area-preserving and thus $\det(\MM)=1$~\cite{meiss2015thirty, smiet2019mapping}.
Therefore, $\MM\in\SL$, the special linear group of order two over the reals. 

$\SL$ is a Lie group whose elements act as area-preserving transformations of the plane, i.e.\ all possible ways (and potentially more) that the Poincar\'e map can behave linearly around a fixed point. 
$\SL$ consists of three subsets\footnote{note these are not sub\emph{groups} as they are not closed under composition} that act in fundamentally different ways:
\begin{itemize}
  \item The \emph{elliptic} subset ($|\Tr(\MM)|<2$): Are conjugate to a rotation. All points lie on invariant sets that are elliptic. It contains the Lie subgroup $K= \left\{\begin{pmatrix} \cos(\theta) & -\sin(\theta) \\ \sin(\theta) & \cos(\theta) \end{pmatrix}, \theta \in[0,2\pi) \right\}$.
    \item The \emph{hyperbolic} subset ($\Tr(\MM)|>2$): Act as a squeeze + expansion  mapping. All points lie on invariant sets that are hyperbolic. It contains the Lie subgroup  $A=\left\{\begin{pmatrix}r & 0 \\ 0 & r^{-1}\end{pmatrix}, r\in\RR_{>0} \right\}$. 
\item The \emph{parabolic} subset ($|\Tr(\MM)|=2$): Act as a shear mapping. All points lie on invariant sets that are straight lines. It contains the Lie subgroup $N=\left\{ \begin{pmatrix} 1 & x \\ 0 & 1 \end{pmatrix} x\in \RR \right\}$ . 
\end{itemize}
According to the Iwasawa decomposition~\cite{iwasawa1949some} every element of $\SL$ can be written as the product of an element of $K$ with an element of $A$ with an element of $N$. 
In other words, every mapping can be understood as the application of a shear, a squeeze and a rotation. 

An \emph{invariant set} of the transformation is a set that is left invariant under the mapping, i.e. maps to itself. 
A fixed point with an elliptic linearized matrix $\MM$ has elliptic invariant sets. 
The magnetic field thus maps such an ellipse to itself, and such a point is often called an o-point. 
Around a surface with rational rotational transform, where all field lines close on themselves, $\MM$ will be parabolic. 
A fixed point around which the invariant sets are hyperbolic is called an x-point. 

We note that the hyperbolic subset consists of two disjoint parts, one where $\Tr(\MM)>2$, and the other where $\Tr(\MM)<-2$.
The eigenvalues of $\MM$ are given by the characteristic equation $\lambda_\pm = (-\Tr(\MM) \pm \sqrt{\Tr(\MM)^2 - 4})/2$. 
When $\Tr(\MM)>+2$, $\MM$ has two real, positive eigenvalues which correspond with the asymptotes of the hyperbolic invariant surfaces.
When $\Tr(\MM)<-2$, the two eigenvalues are both negative. 
Therefore, a point along an eigendirection will be mapped to the \emph{opposite} side of the fixed point. 
Invariant sets are hyperbolic, but each point is mapped to the opposite hyperbolic branch upon application of the map. 
Because of this reflection in the fixed point, such points are called~\emph{reflection hyperbolic}\cite{lichtenberg1992chaotic}. 
Other authors have referred to such points as \emph{hyper-hyperbolic}~\cite{solov1970plasma}, and \emph{alternating-hyperbolic}~\cite{smiet2020bifurcations}, and \emph{M\"obiusian}~\cite{wei2023invariant}. 

Through the Iwasawa decomposition we see that a reflection-hyperbolic mapping can be generated by combining an integer and a half rotation (resulting in the negative identity matrix $-\mathbb{I}$ with trace -2), with a squeeze mapping.
In the linearization of the map generated by $\MM$, the hyperbolic surfaces extend to infinity, but this is rarely the case for the full Poincar\'e section.
A fusion magnetic field is so designed that the bulk of the field line trajectories lie on closed magnetic surfaces. 
An integrable field can smoothly transition from hyperbolic geometry to closed magnetic surfaces if the \emph{critical set}, the trajectories asymptotically departing from the fixed point smoothly connect to the trajectories asymptotically approaching the fixed point forming a~\emph{lemniscate}
\textbf{The surface swept out by this lemniscate, rotated an integer and a half-times before closing on itself, has the topology of an immersed Klein bottle.}

This is not limited to magnetic field line flow, in dynamical systems, a reflection hyperbolic point can be created during a period-doubling bifurcation~\cite{Meyer1992}.
During this bifurcation, an elliptic point becomes a reflection hyperbolic point and two elliptic points that map to each other of double the period on either side. 
The \emph{critical set} is an invariant set that contains a hyperbolic point.
The critical set of the reflection hyperbolic point is a lemniscate. 

The period-doubling bifurcations is often studied purely in the context of its map~\cite{mackay1982renormalisation}. 
But for the magnetic field (and more generally for 1.5D Hamiltonian systems or even general dynamical systems) the map is generated by a 'time'-dependent flow. 
A time-dependent flow of period $T$ can be seen as a series of diffeomorphisms of the plane that maps the invariant sets to themselves at time $T$. 
For a map containing a point that has undergone a period-doubling bifurcation, this flow must take one of the period-doubled points to the other. 
If we assume boundedness of the critical set at all intermediate times, this can only be accomplished by an integer-and-a-half rotation of the critical set, the lemniscate, as it maps to itself.
Therefore, if a period-doubling occurs in a map that can be associated with a time-dependent flow this bifurcation creates a critical set with the topology of an immersed Klein bottle. 

\section{Topological index and vector fields on the Klein bottle\label{sec:index}}
The Hopf-Poincare index theorem relates zeros of a vector field on a manifold $\Ma$ with the topology of that manifold. 
We consider here only vector fields on two-dimensional manifolds.
Let $\Ma$ be a 2 dimensional manifold with $\vv$ a vector field in $T\Ma$ it and $\xx_0$ be an isolated zero of $\vv$. 
Take $\gamma: S^1\rightarrow \Ma$ a loop enclosing that zero. 
The map $g: \xx \mapsto \vv(\xx)/|\vv(\xx)| \in S^1$ sends points to vectors of unit length which we identify with points on the circle.
The restriction of $g$ to the curve $\gamma$ is a map from a circle to a circle: $g|_\gamma: S^1\rightarrow S^1$. 
The \emph{index} of the zero $\xx_0$ equals the degree  of the map $g|_\gamma$, also called the winding number of $\gamma$.

If a loop $\gamma'$ contains more than one zero, then the winding number of $\gamma'$ equals the sum of the indices of the zeros enclosed. 
The Hopf-Poincare index theorem states that the sum of the indices of \emph{any} vector field on $\Ma$ must equal the Euler characteristic $\Ma$
The torus and the Klein bottle are the only two compact two-dimensional manifolds with Euler characteristic 0. 

We will now prove that the `usual' immersion of the Klein bottle cannot occur as the surface spanned by trajectories in nowhere-vanishing fields. 
A Klein bottle embedded in $\RR^3$ must necessarily be self-intersecting. 
If we cut this immersed Klein bottle along the line of self-intersection, then for the `usual` immersion we obtain a disc and a tube with a disc removed. 
Since our field is nowhere-vanishing and tangent to the manifold surface, it must lie along the line of self-intersection which is the boundary of the disc. 
The winding number calculated on the boundary of the disc is therefore $+ 1$, and the disc must contain at least one zero of the vector field, leading to contradiction.

For the lemniscate Klein bottle immersion, the line of self-intersection does not split the surface into two disjoint sections. 
The above contradiction does not occur, and the vector fields we will show below are direct examples of vector fields that contain a magnetic surface with immersed Klein bottle topology. 

There is a third common immersion of the Klein bottle, described by~\citet{lawson1970complete}. 
This surface can be described as a one-parameter family of circles.
The circle of self-intersection of the Klein bottle  does not partition into a disc either,
Whether this immersion can occur as a magnetic surface is unknown. 


\section{Klein bottle magnetic fields\label{sec:realisation}}

\begin{figure}
  \centering
  \begin{minipage}{1.0\linewidth}
    \centering
    \includegraphics[width=\linewidth]{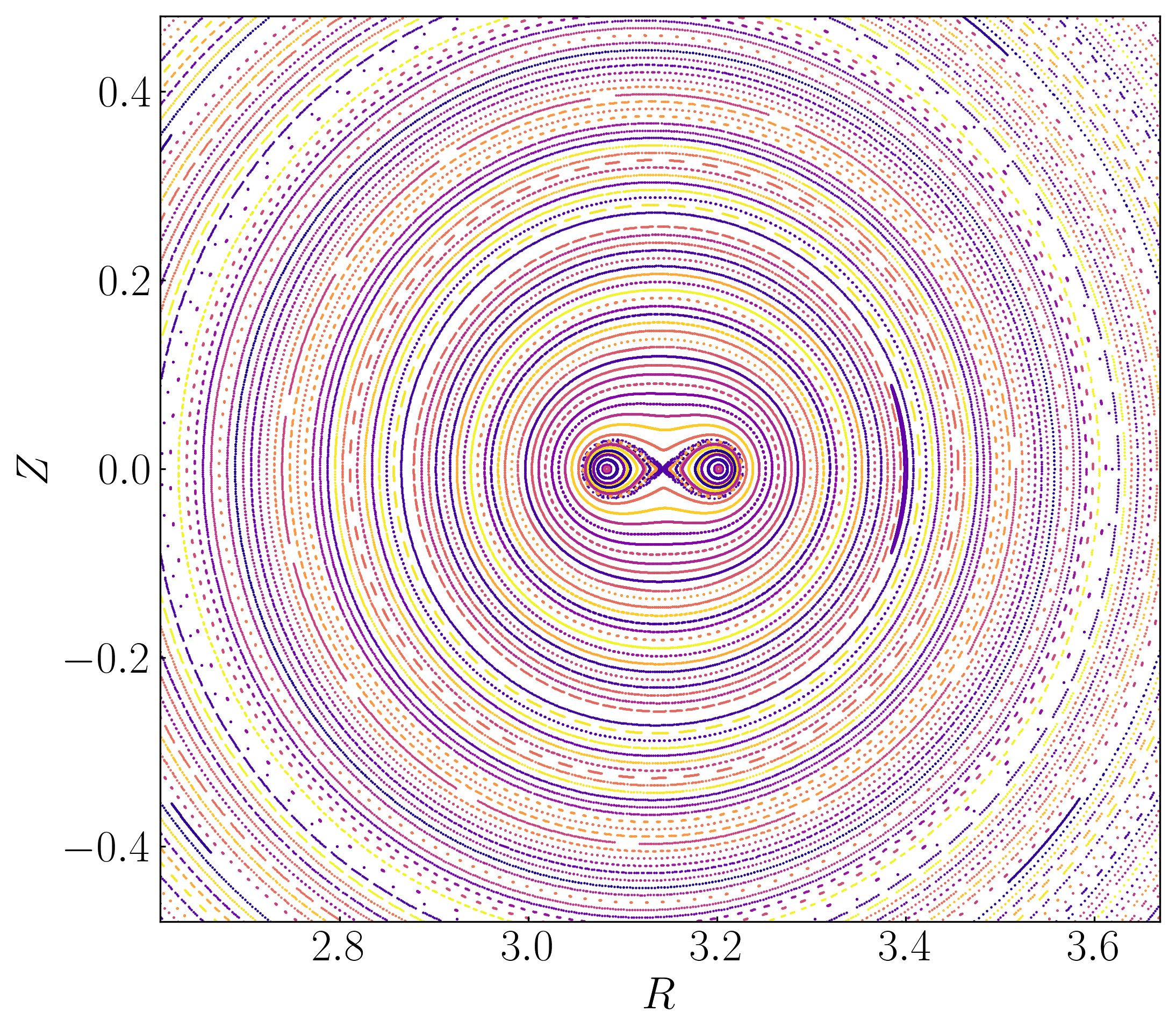}
  \end{minipage}
  \hfill
  \caption{Poincar\'e section of tokamak equilibrium with $q_0=2/3$ ($\imath=1.5$) perturbed with a $2/3$ mode. 
  The axis is reflection-hyperbolic, and its critical set is in the shape of a lemniscate, which is mapped to itself with 1.5 rotations, thus forming a Klein bottle magnetic surface. 
    Adapted from~\cite{smiet2020bifurcations}}\label{fig:althyp}
\end{figure}

In this section we briefly give some examples of vector fields with immersed Klein bottle topology. 
We discuss plasma physics literature in which reflection-hyperbolic points in the magnetic field line map were observed.
We also find examples in the QUASR stellarator database where this geometry occurs. 

The first mention (to the authors knowledge) of reflection hyperbolic points in magnetic fields was by~\citet{solov1970plasma} in their 1970 review. 
They lay out different configurations of magnetic fields in three dimensions using the normal form approach to analyzing bifurcations. 
They specifically note the case of reflection-hyperbolic fixed points, which they label ``hyper-hyperbolic''. 
In their analysis of the normal forms however, they do not make a distinction between regular hyperbolic and reflection hyperbolic fixed points, and they do not note the Klein bottle topology, potentially due to the fact that lemniscate Klein bottle had not been published yet when this review came out. 

Reflection hyperbolic points also play a role in a theoretical model for sawtooth oscillations in tokamak fusion reactors by \citet{smiet2020bifurcations}. 
The sawtooth consists of a periodic sharp drop in core temperature, followed by a slow ramp. 
Most theories assume the crash is caused by an instability occurring when the rotational transform is $\imath\sim 1$, which is often the case, but several observations have also shown that the crash can also occur when the rotational transform is about $1\tfrac{1}{2}$ ($q=\tfrac{3}{2}\approx.7$). 
Ideal stability calculations show that when $\imath=1\tfrac{1}{2}$ there is an unstable mode with mode numbers $n=2, m=3$ that displaces the plasma towards axis in one direction, and away from it in another, generating a squeeze mapping. 

A Poincar\'e section of the field to which a small amplitude of this perturbation was added is shown in figure~\ref{fig:althyp}. 
The axis is reflection-hyperbolic, and the critical set is a lemniscate. 
This lemniscate maps to itself with rotational transform $1\tfrac{1}{2}$, and thus the magnetic surface spanned by these trajectories is an immerssed Klein bottle, though this was not realized at publication. 
When the perturbation amplitude is increased, the field lines around the reflection-hyperbolic fixed point becomes chaotic and the surface with Klein bottle topology disappears.


\begin{figure*}
  \centering
  \begin{minipage}{0.35\linewidth}
    \centering
    \includegraphics[width=\linewidth]{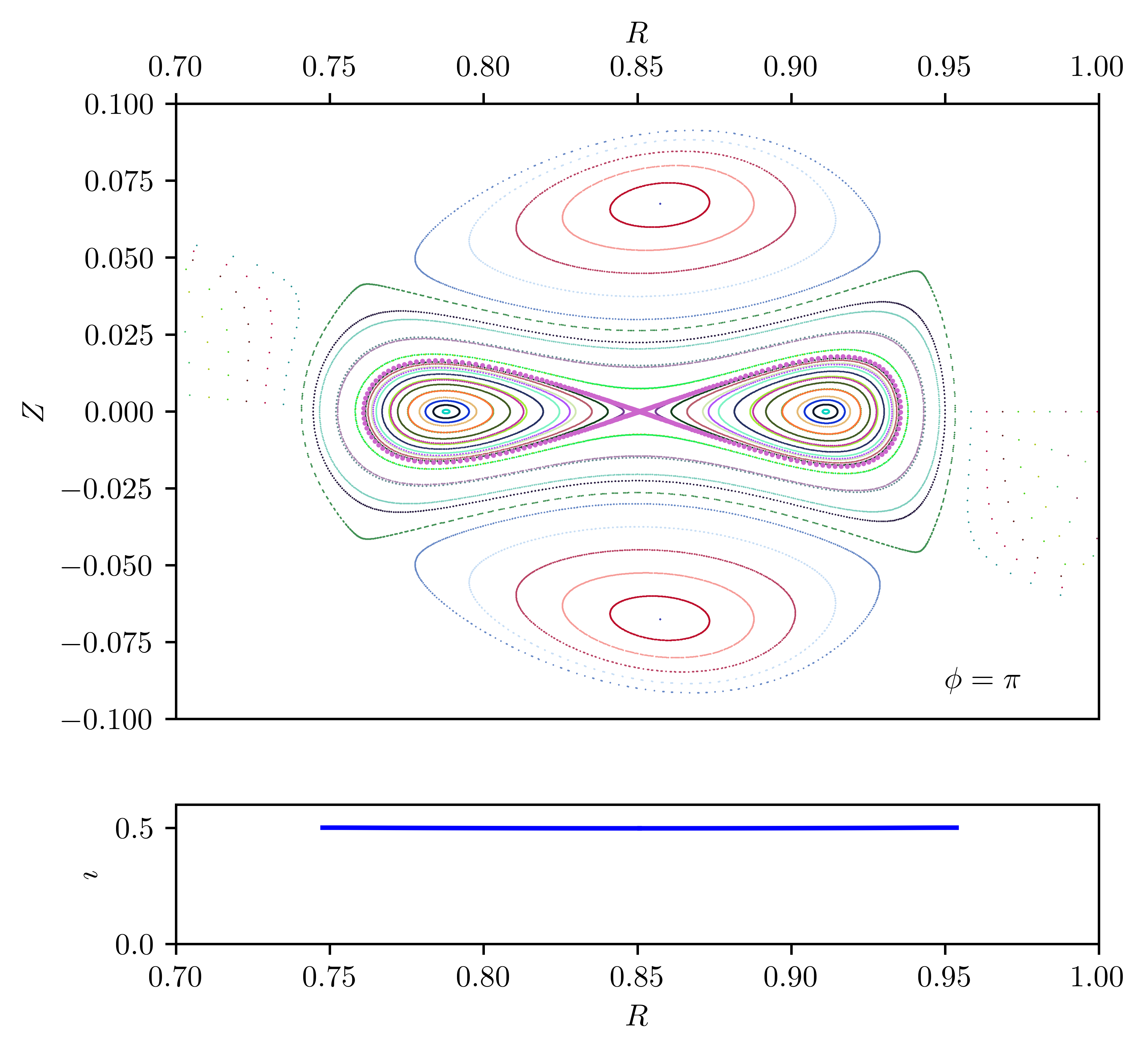}
  \end{minipage}
  \begin{minipage}{0.55\linewidth}
    \centering
    \includegraphics[width=\linewidth]{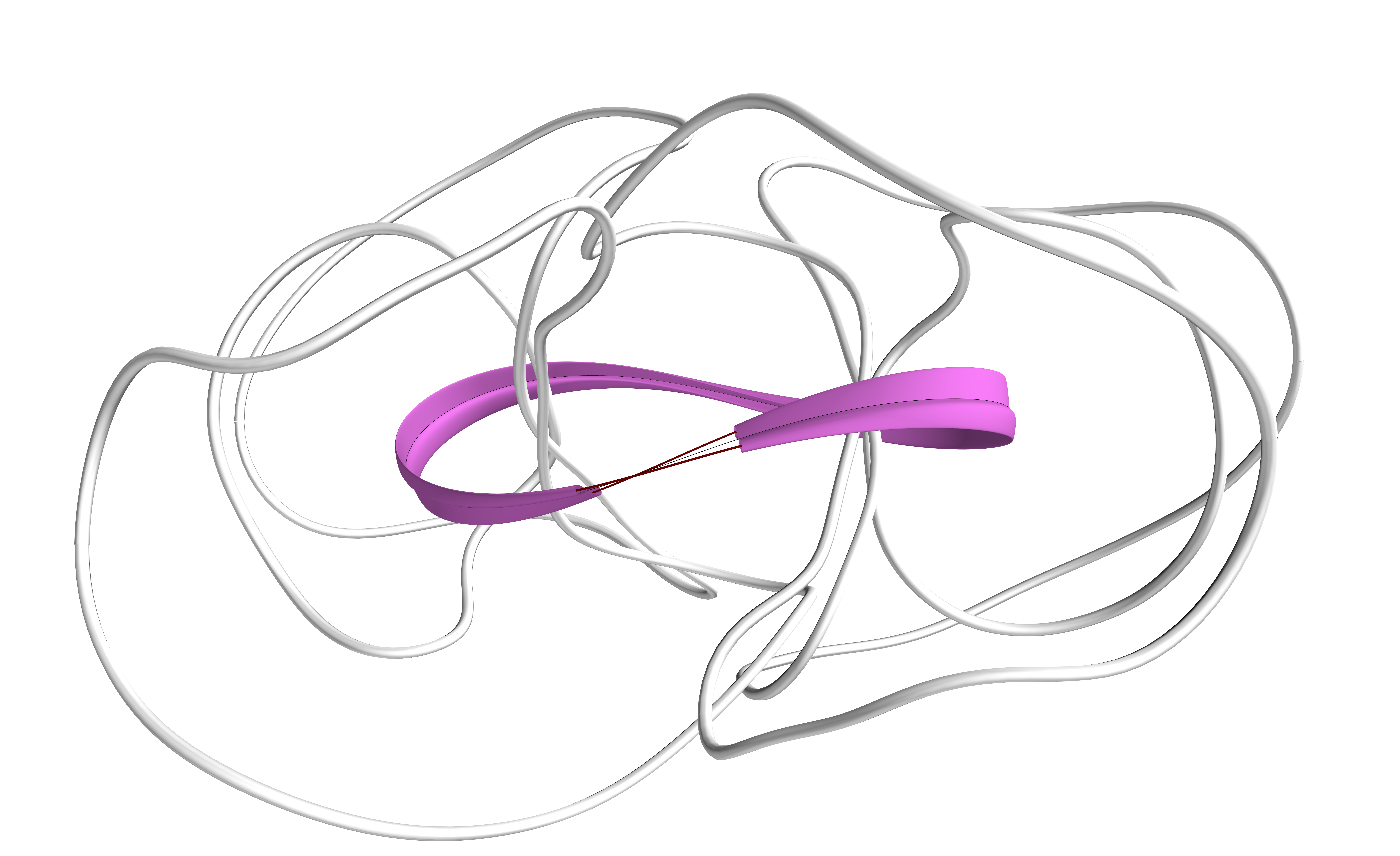}
  \end{minipage}
  \begin{minipage}{\linewidth}
    \centering
    \includegraphics[width=\linewidth]{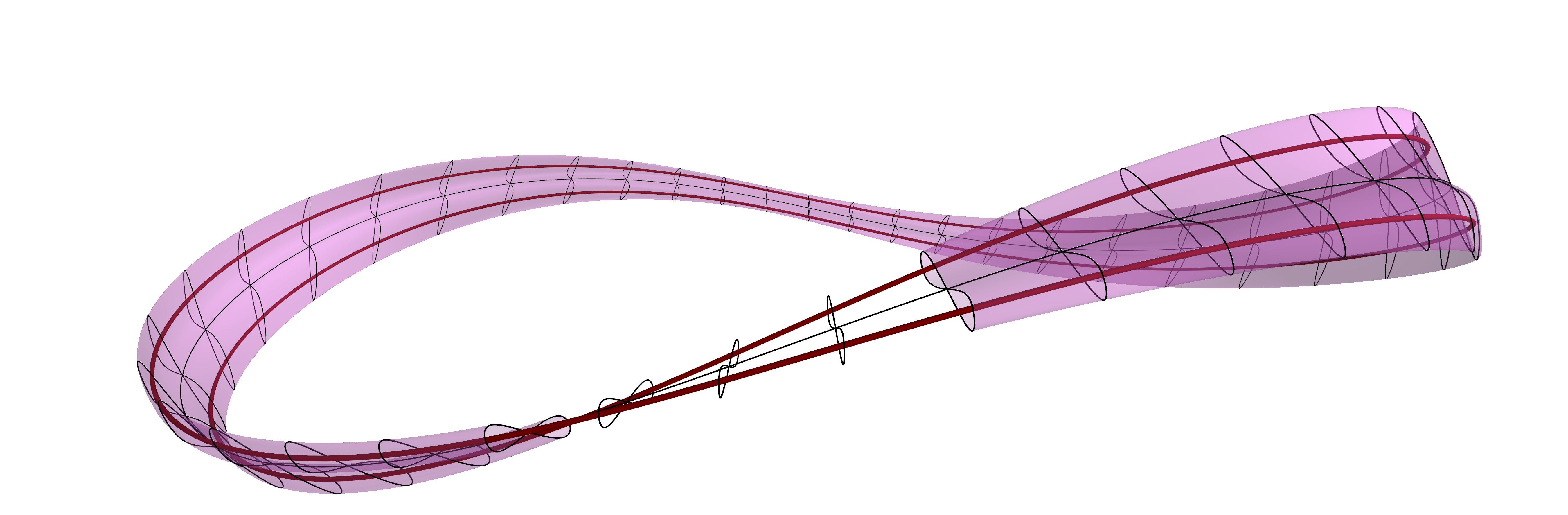}
  \end{minipage}
  \hfill
  \caption{A magnetic surface with Klein bottle topology in QUASR0107534. (top left): Poincar\'e section showing the lemniscate magnetic surface and surrounding trajectories. (top right): The Klein bottle magnetic surface and the filamentary coils generating the magnetic field.  (bottom): zoomed-in view the magnetic surface with Klein bottle topology. The lemniscate cross-section is shown at several constant-$\phi$ planes. The period-doubled trajectory (red) and trajectory of the reflection-hyperbolic point (thin black) are also included. .}\label{fig:quasr}
\end{figure*}

The QUASR database~\cite{giuliani2024comprehensive, giuliani2024direct} also contains stellarators that create magnetic fields which contain surfaces with immersed Klein bottle topology. 
This database contains over 370,000 different optimized quasi-axisymmetric stellarator configurations with coils. 
The optimization is performed on a set of surfaces and coils simultaneously, by matching the surfaces to the magnetic field through minimizing the quadratic flux and simultaneously optimizing the surfaces for a mix of properties including quasisymmetry. 
The match between the surfaces and the field is not always perfect, which is fortunate if one is interested in exploring fields with other topologies~\cite{daies2025topology}.

Especially when the surfaces have close to rational rotational transform, a small perturbation can cause a large variation in the magnetic field~\cite{pedersen2016confirmation}.
As we have seen, magnetic surfaces with immersed Klein bottle topology occur when the rotational transform is integer-and-a half and the field is perturbed so as to cause a period-doubling bifurcation.
We therefore manually search the database for configurations with one field period, rotational transform around 0.5, and large quasisymmetry error. 
A fraction of these configurations contain a reflection-hyperbolic point surrounded by a lemniscate surface. 

We pick one of these configurations, with identifyer \href{https://quasr.flatironinstitute.org/model/0107534}{0107534}. 
First we calculate a Poincar\'e plot by directly integrating the field using Biot-Savart on the coils. 
This is shown in figure~\ref{fig:quasr}. 
The rotational transform $\imath=0.5$, and the critical set of the hyperbolic fixed point is a lemniscate. 

We calculate the 3-dimensional magnetic surface with immersed Klein bottle topology by integrating the field from the points on the lemniscate to 200 different toroidal angles. 
This sweeps out the pink surface shown in~\ref{fig:quasr} (top right, bottom). 
We also locate the reflection hyperbolic point and the period-doubled point and integrate the fields line passing through them.
These are shown as black and the red curve respectively in figure~\ref{fig:quasr}(bottom). 
The surface cross-section is marked with black lines at several cross sections, clearly showing the rotating lemniscate, and hence that this surface has Klein bottle topology.

%
%

\section{Conclusions \label{sec:conclusions}}

Magnetic fields can contain magnetic surfaces with the topology of the lemniscate immersion of the Klein bottle.
Such surfaces can occur when the rotational transform half-integer valued, through a period doubling bifurcation of a fixed point of the Poincar\'e map. 
Examples of this have been shown in stellarator and tokamak fields. 
The magnetic surfaces are of the `lemniscate' type, the`usual` immersion of the Klein bottle cannot occur without points where the field vanishes. 

It should be noted that an immersed Klein bottle is not a Klein bottle because of the self-intersection.
Surfaces with immersed Klein bottle topology in three dimensions are not exclusive to magnetic fields, but could be generated by any dynamical system. 
There is surprisingly little mention of critical sets with this topology in the dynamical systems theory, even though reflection hyperbolic fixed points and period-doubling bifurcations are well studied.
One most often encounters studies of dynamical systems \emph{on} the Klein bottle~\cite{markley1969poincare}. 

The author is a plasma physicist and not an expert on dynamical systems theory, but will attempt a brief literature review.
In four dimensions (such as 2 degree of freedom Hamiltonian systems), true Klein bottles can appear. 
~\citet{bolsinov2004integrable} discusses the classification of 2 degree of freedom integrable Hamiltonian systems $Q$. 
They mention that only circles, tori and Klein bottles and can exist as critical submanifolds of the integral $f$ on $Q$. 
The topological classification of integrable systems with critical Klein bottles needs a different treatment from those which do not~\cite{topalov1994inclusion}. 
The review by Arnol'd~\cite{arnold2013dynamical} discuses bifurcations in general dynamical systems, and it is mentioned that the nontrivial limit sets born from nonhyperbolic cycles with multiplier +1 can have the topology of Klein bottles.

\citet{osinga2003nonorientable} studies nonorientable manifolds in three-dimensional vector fields. 
They show that the manifolds (akin to critical sets, the trajectories that asymptotically approach and leave a closed trajectory) during period-doubling bifurcations, twisted homoclinic bifurcations and saddle-node bifurcations on a limit cycle can be nonorientable.  
The identification with Klein bottles is avoided as it is specifically noted that a Klein bottle can only be embedded in a space of at least four. 
This paper shows that nevertheless,  the immersed Klein bottle can occur in three-dimensional vector fields.
Studies in four dimensions mention invariant sets with Klein bottle topology~\cite{MINDLIN1997177}. 

Trajectories with Klein bottle topology have been observed in other branches of physics. 
For example the laws of motion are Hamiltonian, and certain orbits surrounding irregularly shaped asteroids (3 degree-of-freedom Hamiltonians with 6-dimensional phase space) can have Klein bottle topology~\cite{jiang2015topological}.

Period doubling is often associated with a transition from an integrable state to chaotic dynamics~\cite{mackay1982renormalisation}. 
The 
The magnetic surfaces with Klein bottle topology disappear when the field becomes chaotic. 
When this happens, the reflection hyerbolic point is surrounded by chaotic field lines. 
One can quantify this chaos using methods such as the turnstile area~\cite{smiet2025turnstile, meiss2015thirty, mackay1984transport}, which calculate the phase space area that enter and exit a resonance zone. 
To my knowledge there is little research that specifically adresses the difference between chaotic dynamics around reflection hyperbolic points and regular hyperbolic points (I am only aware of~\cite{miguel2022escape, stirling1998dynamical, sterling1999homoclinic}), though I expect the topology to make a difference. 
Reflection-hyperbolic points are seen in the chaotic regions surrounding stellarator configurations~\cite{daies2025topology}, and could be used in divertors.

In this work it has been emphasized that the field is nowhere vanishing. 
In magnetic fields that are allowed to vanish, magnetic surfaces with other topologies are possible. 
\citet{smiet2017magnetic} show that in the resistive evolution of certain twisted and knotted fields, which contain lines where the field vanishes, magnetic surfaces can exist that foliate a region of space and span three-tori ($\chi=-4$) or even higher genus surfaces. 

The results in this paper will not solve fundamental problems towards the realization of fusion, but they do highlight surprising connections between disparate branches of physics and mathematics. 
It can give us insight into certain instabilities, and warn us to avoid certain configurations with integer-and-a-half rotational transform. 
Understanding the surprising ways in which magnetic fields can behave, and exploring the deep connection with Hamiltonian and general dynamical systems, allows us to exercise our imagination and develop the mathematical tools necessary for making fusion a reality.

\begin{acknowledgments}
I would like to thank Jim Meiss and Robert MacKay for their encouragement, interesting discussions, and correction of this manuscript. 
This work was supported by a grant from the Simons Foundation\ (1013657, JL). 
This work has been carried out within the framework of the EUROfusion Consortium, via the Euratom Research and Training Programme (Grant Agreement No 101052200 — EUROfusion) and funded by the Swiss State Secretariat for Education, Research and Innovation (SERI). Views and opinions expressed are however those of the author(s) only and do not necessarily reflect those of the European Union, the European Commission, or SERI. Neither the European Union nor the European Commission nor SERI can be held responsible for them. 
\end{acknowledgments}

\bibliography{references}
\end{document}